\documentclass[10pt]{article}

\usepackage[english]{babel}
\usepackage{times}

\usepackage[letterpaper,top=0.5in,bottom=0.5in,left=1in,right=1in,headheight=30pt,includeheadfoot]{geometry}
\usepackage{comment}
\usepackage{amsmath,amsfonts}
\usepackage{graphicx}
\usepackage{subcaption}
\usepackage{multirow}
\usepackage{makecell}

\usepackage[colorlinks=true, allcolors=blue]{hyperref}
\usepackage{xcolor}
\usepackage{soul} 
\usepackage{listings}
\usepackage{matlab-prettifier}
\usepackage[title]{appendix}
\usepackage{authblk}
\usepackage{setspace}
\usepackage{lastpage}
\usepackage{fancyhdr}
\makeatletter
\let\ps@plain\ps@fancy
\renewcommand{\@maketitle}{%
  \parindent=0pt%
  \begin{center}
  \@title
  \end{center}
  \par%
  \@author
}
\makeatother

\usepackage{titlesec}
\titleformat*{\section}{\normalsize\bfseries}
\titleformat*{\subsection}{\normalsize\itshape}
\titleformat*{\subsubsection}{\normalsize\itshape}
\titlespacing{\section}{0pt}{*4}{*0}
\titlespacing{\subsection}{0pt}{*4}{*0}
\titlespacing{\subsubsection}{0pt}{*4}{*0}

\usepackage{nomencl}
\makenomenclature

\usepackage{acronym} 

\renewcommand{\d}[1]{\operatorname{d}\!#1} 
\newcommand{\xp}[1]{\operatorname{E}\left[#1\right]}  

\title{\doublespacing \normalsize \textbf{Inference on the Miss Distance in a Conjunction}}

\author[a*]{\normalsize \textbf{\hskip 0.58in J.~Russell~Carpenter}}
\affil[a]{\textit{Space Science Mission Operations, NASA Goddard Space Flight Center, USA, russell.carpenter@nasa.gov}}
\author[b]{\normalsize \textbf{Anthony~C.~Davison}}
\affil[b]{\textit{Institute of Mathematics, Ecole Polytechnique Fédérale de Lausanne, Switzerland, anthony.davison@epfl.ch}}
\author[c]{\normalsize \textbf{Soumaya~Elkantassi}}
\affil[c]{\textit{Department of Operations, University of Lausanne, Switzerland, soumaya.elkantassi@unil.ch}}
\author[d]{\normalsize \textbf{Matthew~D.~Hejduk}}
\affil[d]{\textit{The Aerospace Corporation, USA, matthew.d.hejduk@aero.org}}
\affil[*]{Corresponding Author}

\date{}

\def\frac#1#2{{\textstyle{#1\over#2}}}

\def\tr{{\rm tr}}

\newcommand{\D}[1]{\mathrm{d}{#1}}

\def\bi{\begin{itemize}}
\def\ei{\end{itemize}}
\def\bd{\begin{description}}
\def\ed{\end{description}}
\def\ben{\begin{enumerate}}
\def\een{\end{enumerate}}
\def\bv{\begin{verbatim}}
\def\ev\end{verbatim}

\def\calN{{\mathcal N}}
\def\N{{\calN}}

\def\d{{\delta}}

\def\v{{\varepsilon}}

\def\T{{ \mathrm{\scriptscriptstyle T} }}

\DeclareMathOperator*{\E}{E}
\DeclareMathOperator*{\var}{var}

\def\2to{{\ {\buildrel 2\over \longrightarrow}\ }}

\def\Deq{{\ {\buildrel {\rm D}\over =}\ }}

\def\I1ton{{$I_1,\ldots,I_n$}}
\def\X1ton{{$X_1,\ldots,X_n$}}
\def\Y1ton{{$Y_1,\ldots,Y_n$}}
\def\Z1ton{{$Z_1,\ldots,Z_n$}}
\def\R1ton{{$R_1,\ldots,R_n$}}
\def\e1ton{{$e_1,\ldots,e_n$}}
\def\t1ton{{$t_1,\ldots,t_n$}}
\def\x1ton{{$x_1,\ldots,x_n$}}
\def\y1ton{{$y_1,\ldots,y_n$}}
\def\z1ton{{$z_1,\ldots,z_n$}}

\begin{document}
\maketitle
\vskip 12pt

\textbf{\centerline{Abstract}}\\
\indent Over the last quarter-century, spacecraft conjunction assessment has focused on a quantity associated by its advocates with collision probability. This quantity has a well-known dilution feature, where it is small when uncertainty is large, giving rise to false confidence that a conjunction is safe when it is not. An alternative approach to conjunction assessment is to assess the missed detection probability that the best available information indicates the conjunction to be safe, when it is actually unsafe. In other words, the alternative seeks to answer the question of whether unknowable errors in the best available data might be especially unlucky. A proper implementation of this alternative avoids dilution and false confidence. Implementations of the alternative use either significance probabilities (p-values) associated with a null hypothesis that the miss distance is small, or confidence intervals on the miss distance. Both approaches rely on maximum likelihood principles to deal with nuisance variables, rather than marginalization. This paper discusses the problems with the traditional approach, and summarizes other work that developed the alternative approach. The paper presents examples of application of the alternatives using data from actual conjunctions experienced in operations, including synthetic scaling to highlight contrasts between the alternative and the traditional approach. 


\printnomenclature

\section*{Acronyms/Abbreviations}
\begin{acronym}[NASA] 
    \acro{CA}{Conjunction Assessment}
    \acro{CDM}{Conjunction Data Message}
    \acro{CI}{Confidence Interval}
    \acro{DR}{Detection Rate}
	\acro{FAR}{False Alarm Rate}
	\acro{HBR}{Hard Body Radius}
	\acro{MDR}{Missed Detection Rate}
	\acro{PDF}{Probability Density Function}
    \acro{ROC}{Receiver Operating Characteristic}
	\acro{TCA}{Time of Closest Approach}
\end{acronym}

\section{Introduction}

The decision whether to take action to mitigate the risk of a collision of two objects in space involves a great deal of uncertainty.  It often reduces to considering whether a region associated with uncertainty in their relative positions overlaps a region representing the combined sizes of the objects. This assessment can often be simplified by projecting these regions into a plane normal to the relative velocity vector at the \ac{TCA}. Although many would trace the origins of this problem to Kessler’s work in the late 1970s predicting the ``creation of a debris belt’’ \cite{KellerDebrisBelt,KESSLER198139}, the closely related problem of accounting for uncertainty when planning a planetary flyby appears at least a decade earlier in internal NASA reports concerning the planning for interplanetary missions to Mars and Venus. In these early works, it appears that mission designers planned to bias the trajectory to avoid overlap between a disk related to the planetary target and an elliptical region derived from orbit determination uncertainties. These two-dimensional constructs result from projection of their three-dimensional counterparts into a plane normal to the hyperbolic trajectory asymptote at the time of closest approach.  In current practice, consideration of such decisions is typically guided by a quantity, $p_c$, that is intended to represent a probability of collision~\cite{Fost92:collp,Akel00:collp,Coppo:12velPc,Hall:2000aa,hall21:colrate} and treats the uncertainty region as a \ac{PDF}. The mean is given by the estimated relative state, predicted forward to the \ac{TCA}. The shape of the \ac{PDF} is governed by a covariance matrix derived via the orbit determination process from errors assumed to be present in the tracking data\footnote{Sometimes the process of generating the predicted covariance also includes priors associated with parameters affecting the prediction process, and may even include the addition of covariances representing fictitious ``process noise’’ effects.}. A similar process is used when predicting the risks posed to Earth by asteroids. A recent occurrence involving the asteroid 2024~YR4\cite{wasser2025latest} illustrates a feature of the $p_c$-based approach to risk assessment, which some have called ``dilution’’. \cite{Alfano:2005aa}: when the uncertainty is large in comparison to the miss distance,  $p_c$ will be small.  Initially, as more observations are collected, the uncertainty begins to shrink and $p_c$ begins to rise, sometimes alarmingly. Then, with further reduction in the uncertainty, $p_c$ will often decline again. Unless it doesn’t. In another recent occurrence, the operational but non-maneuverable TIMED and the defunct COSMOS 2221 satellites had a close approach during which $p_c$ continued rising to high levels all the way up to the \ac{TCA} \cite{foust2024nasa}, although no collision occurred. Operators have grown used to dealing with the features and quirks of $p_c$-based conjunction risk assessments and generally deal with these by planning risk mitigation activities well in advance of the \ac{TCA}, often when the uncertainties remain large, such that the vast majority of such mitigations end up as false alarms. Apart from those who have conducted destructive weapons testing in low Earth orbit, all actors agree that a ``Kessler syndrome’’ would have disastrous consequences for space development, but false alarms come at a cost to satellite operations. No decision under uncertainty can be ideal, but one hopes that being conservative, e.g., by setting a $p_c$ decision threshold at a small value such as $10^{-4}$,  should adequately limit the fraction of wrong decisions.

It may have come as a surprise to the operations community when two recent papers described different and more severe concerns with basing space object collision mitigation on $p_c$. One of these papers~\cite{Balch2019SatelliteCA} evocatively labeled this concern ``false confidence’’.  The critique is essentially that $p_c$ can fail to rise above any particular decision threshold when a collision will actually occur. In other words, the missed detection rate of a $p_c$-based decision procedure can be orders of magnitude higher than one might imagine it to be, especially if one imagines that the missed detection rate has something to do with the decision threshold. A simple example in the sequel demonstrates this point explicitly. The solution proposed by \cite{Balch2019SatelliteCA} was to reinterpret the uncertainty regions as confidence regions, and revert to use of confidence region overlap as a decision criteria. The other of these papers~\cite{Elkan22:StatFormCA}, on which the present work is based, described a somewhat more precise approach, involving either the construction of confidence intervals on the miss distance, or equivalently significance probabilities, also known as $P$-values. While an earlier article~\cite{Carpenter:2019aa} had similarly proposed credible sets on the miss distance as a slight improvement over $p_c$, \cite{Elkan22:StatFormCA} derived the confidence interval in a more rigorous manner, and hence provided a marked improvement in the error characteristics of any resulting decisions.

The primary contributions of the present paper are to further develop the ideas of~\cite{Elkan22:StatFormCA}. It provides a simple example illustrating the deficient missed detection properties of $p_c$. It compares and contrasts methods of computing confidence intervals and their connections to $p_c$ and $P$-values. It describes an alternative computation to that proposed in \cite{Elkan22:StatFormCA} which appears to be more robust to poorly conditioned covariance data that often occurs in operations. It discusses a manner of choosing decision thresholds based on prior densities. It contributes a computation of a truncated uniform prior that supports the adequacy of the more-often assumed infinite uniform prior as a step toward Bayesian inference. It describes a method to derive empirical Bayesian priors from prior conjunction data. It provides \ac{ROC} curves that compare $p_c$- and $P$-value-based decision procedures, and which show that the latter dominate the former. Finally it offers conclusions on the path forward for incorporating statistically sound inference on the miss distance, via the computations described herein, into the space operations community.

\section{Problem Statement, Assumptions, and Methods}

\subsection{Space Object Conjunction Assessment}\label{sec:ca}
Most space object conjunctions occur over very short time intervals at very high relative velocities. These conditions permit several simplifying assumptions that reduce the geometry to two dimensions, namely, within a short interval around the time of closest approach: (1) there is only one close approach; (2) relative motion occurs along a straight line; (3) velocity uncertainty may be neglected. These assumptions imply that the conjunction need only be considered in a plane normal to the relative velocity at the \ac{TCA}. Additionally, operators often assume that no unknown perturbations occur between the orbit determination solution epoch and the \ac{TCA}. 

\begin{figure}[!htb]
	\centering
	\includegraphics[width=0.5\linewidth]{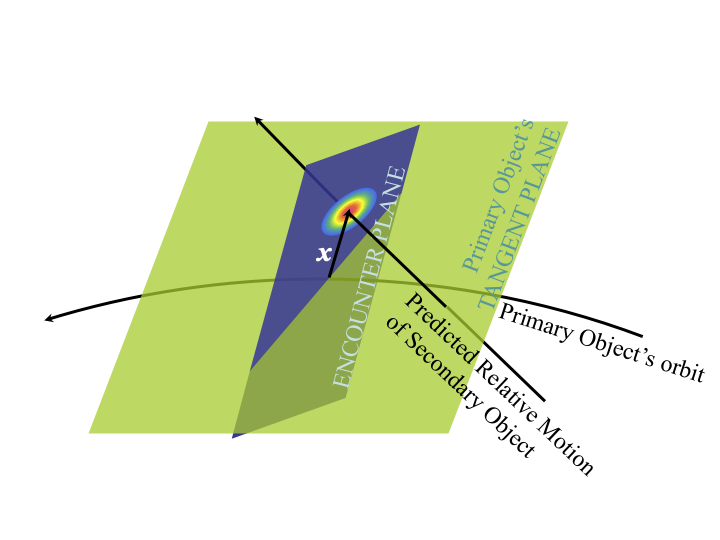}
	\caption{\label{fig:encpl}Geometry of a space object conjunction.}
\end{figure}

Figure~\ref{fig:encpl} depicts the geometry of a such a close approach between two space objects. The tangent plane, i.e., the local horizontal plane of the primary object's orbit, is shown for reference. The encounter plane, also known as the conjunction plane, is normal to the relative velocity vector. The encounter plane projection of the predicted relative position vector, $x$, has a confidence/credible region indicated by the colored ellipses, with warmer colors indicating higher values. The true relative position vector at closest approach, $\xi$, is unknown. Not shown is the \ac{HBR} sphere, enclosing the combined size of the two objects. When a coordinate frame is attached to the encounter plane, the center of the \ac{HBR} serves as the origin.

When one or both of the objects involved in the conjunction is an active satellite, then the situation depicted in  Figure~\ref{fig:encpl} faces an operator with the decision whether to take some action to mitigate the risk of a collision between them. It is tempting to view the multi-colored ellipses in Figure~\ref{fig:encpl} as contours of a \ac{PDF} $f(x;\xi)$, for the errors associated with the prediction. However, since the ellipses are centered on $x$, the predicted relative position, and not $\xi$, the true relative position, this is not a correct interpretation, even if the covariance matrix from which these ellipses is derived is correct. Rather, any particular elliptical region centered on the predicted relative position either will, or will not, contain the true relative position when the \ac{TCA} arrives. This suggests two interpretations of the multi-colored ellipses in Figure~\ref{fig:encpl}.  

One interpretation of such regions is to imagine many hypothetical repetitions of the conjunction that differ only due to errors in the tracking data underlying the predictions. Here the true relative position, $\xi$, is regarded as an unknown, but fixed, quantity, and probability relates to the repetitions, under which some fraction of the elliptical regions will contain $\xi$ at the \ac{TCA}. If one imagines the number of repetitions growing ever larger, this fraction will stabilize, such that in 99\% of the repetitions, for example, the light blue contour will contain $\xi$. But this says nothing about whether any particular repetition has a region of any particular color that contains $\xi$. From this perspective all probabilities are computed with respect to these repetitions of the data, and are then transformed into so-called ``confidence statements'' about regions that may contain the unknown $\xi$.  

It is tempting to use the ``approximation'' $f(x;\xi)\approx f(x';x)$, i.e., the predicted relative position is a ``good enough'' or ``best available'' approximation of the true relative position, and then to integrate $f(x';x)$ over the \ac{HBR}, thereby obtaining an estimated probability of collision 
\begin{equation}
	\hat{p}_c = \int_{x'\in\text{\ac{HBR}}} f(x';x)\,\D{x'}.
\end{equation}
But even if by pure chance $x=\xi$, 
\begin{equation}
	p_c = \int_{x'\in\text{\ac{HBR}}} f(x';\xi)\,\D{ x'}
	\label{eq:$p_c$}
\end{equation}
gives the probability that the noisy prediction, $x$, will pass within the \ac{HBR}, rather than the probability that the true miss vector, $\xi$, lies inside the \ac{HBR}. 

The second interpretation dispenses with the need to envisage large numbers of hypothetical repetitions, but at the cost of abandoning the association of probability with relative frequencies. Instead, one chooses, or is given, a prior \ac{PDF} for the true relative position, $\xi$. With the prior, $f(\xi)$, one may treat $\xi$ as a random variable, and in principle compute its conditional density, given the prediction $x$, $f(\xi\mid x)$, using Bayes' Rule, as
\begin{equation}
	f(\xi\mid x) = \dfrac{f(x;\xi)f(\xi)}{\int f(x;\xi)f(\xi)\,\D{\xi}} = \dfrac{f(x\mid \xi)f(\xi)}{\int f(x\mid \xi)f(\xi)\,\D{\xi}} = \dfrac{f(x\mid \xi)f(\xi)}{f(x)}.
	\label{eq:condensty}
\end{equation}
\nomenclature{$f(x;\xi)$}{probability density of a prediction $x$, centered on the unknown non-random parameter $\xi$;  if $\xi$ is treated as if it were a random variable, $f(x;\xi)=f(x\mid\xi)$}
\nomenclature{$f(\xi)$}{prior density for $\xi$, treated as a random variable}
\nomenclature{$f(\xi\mid x)$}{conditional density for the parameter $\xi$, treated as if it were a random variable, given a realized value of $x$}
\nomenclature{$x$}{projection of a prediction of the relative position vector into the encounter plane, which contains random variations due to tracking errors}
\nomenclature{$\xi$}{projection of the non-random, but unknown, true relative position vector into the encounter plane}

\noindent This interpretation relies on the provision of a prior density on the truth, $f(\xi)$, rather than solely on observations of the world, so it could be criticized as less than fully objective. The analogue of a confidence region in this setup is often called a ``credible region'', to indicate that it relies upon a different interpretation of probability. The use of Bayes' Rule to update the prior \ac{PDF} $f(\xi)$ to the so-called posterior \ac{PDF} $f(\xi\mid x)$ given in Eqn.~\eqref{eq:condensty} has led to this interpretation being known as Bayesian.

The Bayesian posterior probability that the true position $\xi$ lies within the \ac{HBR}, 
\begin{equation}
	\int_{\xi\in\text{\ac{HBR}}} f(\xi\mid x)\, \D{\xi},
\end{equation}
requires the specification of a prior \ac{PDF} so that the true miss vector $\xi$ can be treated as a random variable, and one might better term it a ``credibility of collision''.  Clearly its value depends on the chosen prior $f(\xi)$.  If we assume that this is uniform, with a constant value over of all of infinite space, and if one is careful with limits, and if $f(\xi\mid x) = f(x\mid \xi)$ for all $\xi$ and $x$, as with a Gaussian \ac{PDF}, then it can be seen that  Eq.~\ref{eq:$p_c$} also represents a credibility of collision.   This requires not only the uniform prior but also the symmetry leading to the equality $f(\xi\mid x) = f(x\mid \xi)$, so it can be regarded as accidental.  Although a uniform prior on an infinite set is improper, this is not a practical problem unless the posterior density is also improper. 
Section~\ref{trunc.sec}~introduces a truncation of the uniform prior to a more reasonable volume of space, corresponding to the screening volume that was used to identify the existence of the conjunction in the first place, and Section~\ref{EB.sec}~outlines how a prior density might be estimated from data on past conjunctions. 

\begin{figure}[!htb]
	\centering
	\includegraphics[width=0.67\linewidth]{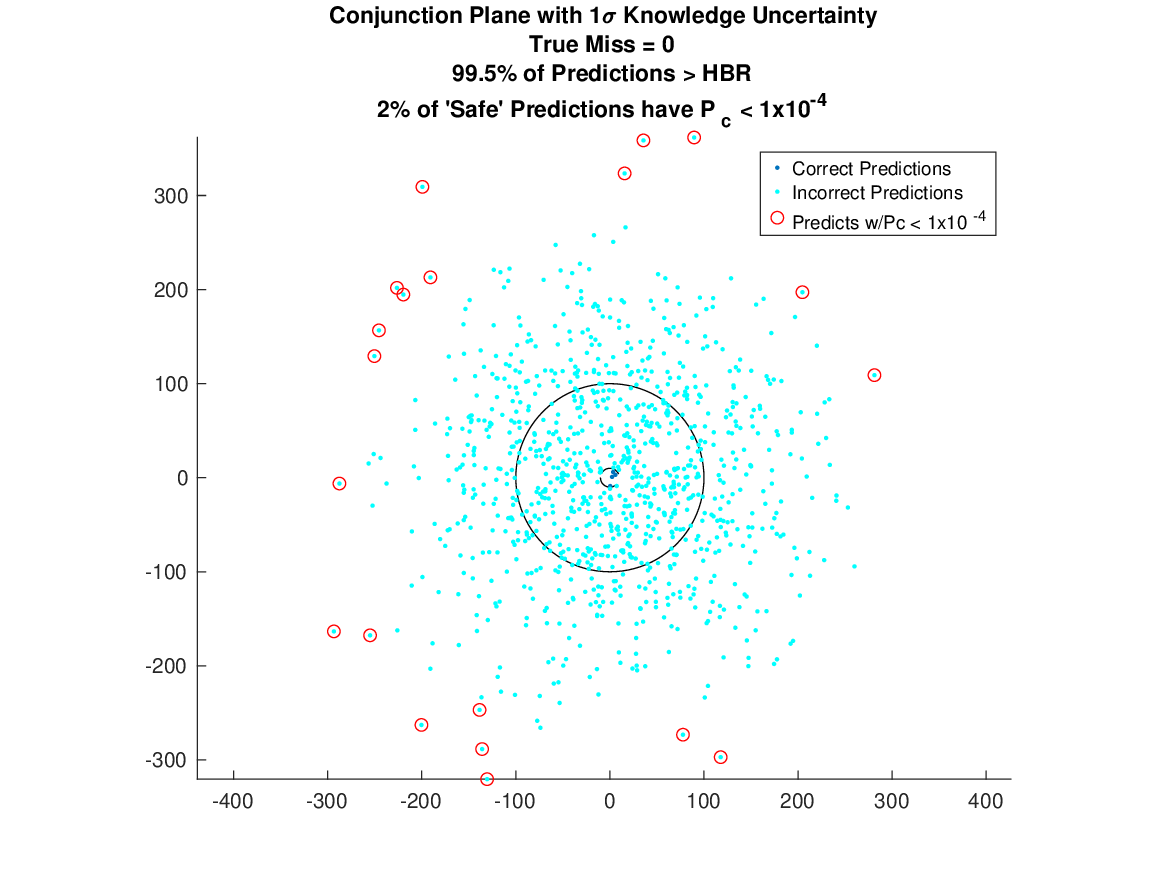}
	\caption{\label{fig:PcZeroMiss} Example with zero true miss distance. The small inner circle represents the \ac{HBR}, while the large outer circle represents the $1\sigma$ error of predictions of the miss vector.}
\end{figure}

Another issue with the computation of $\hat{p}_c$ is often  significant: because the estimated miss distance will tend to be greater than the true miss distance, $\hat{p}_c$ is typically smaller than $p_c$. To see this, suppose that $f(x;\xi)$ is a Gaussian \ac{PDF} with true mean $\xi$ and covariance matrix $D$, and that without loss of generality the  coordinate system in the conjunction plane has been chosen so that $D$ is diagonal, with diagonal entries $d_1^2$ and $d_2^2$.  Then, the expected length of $x$,
\begin{equation}
\xp{\|x\|^2} = \xp{x_1^2 + x_2^2} = \xp{x_1^2} + \xp{x_2^2} = \xi_1^2 + \xi_2^2 + d_1^2 + d_2^2 = \|\xi\|^2+ \tr(D),
\end{equation} 
is strictly greater than $\|\xi\|^2$. 
Thus, a \ac{PDF} for the miss vector centered on $x$ will tend to be further from the \ac{HBR} than a \ac{PDF} centered on $\xi$, and hence $\hat{p}_c$ will tend to underestimate	$p_c$.  Figure~2 of Reference~\cite{Elkan22:StatFormCA} summarizes a study that comprehensively illustrates this phenomenon over a range of samples drawn from cases in which the miss distance is both greater and less than the \ac{HBR}.  Figure~\ref{fig:PcZeroMiss} depicts an even clearer example, where the true miss distance is precisely zero, the worst possible situation. For this example, $f(x;\xi)$ is Gaussian with $\sigma_1 = \sigma_2 = 100~\text{m}$, and the radius of the combined hard-body is $10~\text{m}$. Of the 10,000~predictions drawn from $f(x;\xi\!=\!0)$, $99.5\%$ have miss distances greater than the \ac{HBR}. Each prediction $x$ has its own value of $\hat{p}_c$, computed with a Gaussian distribution centered at $x$, as would occur in operations.  As the figure shows, $2\%$ of the 10,000 predictions, those circled in red, have values of $\hat{p}_c$ below $1\times 10^{-4}$, which is a common threshold for dismissal in operational settings.  Thus, if this threshold had been chosen to represent one chance in 10,000 of wrongly concluding that a conjunction was safe, the perception of risk in this example would differ appreciably from the reality.

\subsection{Statistical Inference}\label{sec:infernc}

A key aspect of statistical modeling is a clear distinction between observed quantities and the unknowns about which an inference is required.  This is well-illustrated in Figure~\ref{fig:PcZeroMiss}: an operator would know the position of the primary object and its \ac{HBR}, observe just one predicted position $x$ for the secondary object, and hope to use this single $x$ for inference on the true but unknown intersection, $\xi$.  This informs a decision about whether mitigation to avoid a collision is required; we shall discuss this in~\S\ref{sec:comparison}.   Assumptions about hypothetical repetitions of the sampling or about prior knowledge are made only to base an inference about the true but unknown position $\xi$ on an estimated position $x$. 

Under the additional assumption that the \ac{PDF} $f(x;\xi)$ is Gaussian with known covariance matrix $D$, exact inference on $\xi$ is possible.  Without loss of generality the covariance matrix can be taken to be diagonal, and then 
\begin{equation}\label{gaussian-pdf.eq}
f(x;\xi) = \dfrac{1}{2\pi d_1d_2}\exp\left\{- \dfrac{(x_1-\xi_1)^2}{2d_1^2} -\dfrac{(x_2-\xi_2)^2}{2d_2^2}  \right\},\quad -\infty< x_1,x_2,\xi_1,\xi_2<\infty,\quad d_1,d_2>0.
\end{equation}
It follows that under hypothetical repetitions of the conjunction, the squared Mahalanobis distance $\Delta(\xi)=(x_1-\xi_1)^2/d_1^2 + (x_2-\xi_2)^2/d_2^2$ has a chi-square distribution with two degrees of freedom, or, equivalently, $\Delta(\xi)/2$ has a unit exponential distribution.  Such a quantity, which depends on both the data and the unknown quantity of interest and has a known distribution, is called a pivot; it can be used to construct confidence regions, which contain the unknowns with specified probabilities.  In this case the distribution of the pivot is known, and it turns out that the elliptical set 
$$
\mathcal{E}_{1-\alpha} = \left\{ \xi: \Delta(\xi) \leq -2\log\alpha\right\}
$$
centered on $x$ is a $(1-\alpha)\times 100\%$ confidence region for the unknown $\xi$.  If the Gaussian model is correct, then $\mathcal{E}_{1-\alpha}$ contains the true $\xi$ with probability exactly $1-\alpha$ under hypothetical repetitions of the conjunction.  With $\alpha=10^{-4}$, for example, any values of $\xi$ outside the ellipse given by $\Delta(\xi) = -2\log 10^{-4} = 18.42$ would be regarded as incompatible with the data.  The exact distribution also allows a test of the hypothesis that $\xi$ takes a specific value, and in particular that $\xi=0$, i.e., that the true conjunction will be at the center of the \ac{HBR}.  Under this hypothesis, $\Delta(0)/2$ is a unit exponential variable, and thus the so-called $P$-value, the probability $\exp\{ - \Delta(0)/2\}$ that such a variable exceeds $\Delta(0)/2$, is regarded as evidence against this hypothesis, with small values casting greater doubt on it.  Here $\xi=0$ lies outside $\mathcal{E}_{1-\alpha}$ precisely when $\exp\{ - \Delta(0)/2\}<\alpha$.  A test of the hypothesis that $\xi$ lies within the \ac{HBR} would be rejected only if $\max_{\xi\in\ac{HBR}} \exp\{ - \Delta(\xi)/2\}$ was smaller than a suitable value, or equivalently, if $\min_{\xi\in \ac{HBR}} \Delta(\xi)$ was larger than $-2\log \alpha$ for some suitable probability $\alpha$.  

It may be more convenient for an operator to use the $P$-value to screen conjunctions but to examine the ellipse only if this probability exceeds some pre-chosen level.

If the \ac{HBR} is appreciable 
then it may be preferable to test for $\xi\in\ac{HBR}$ rather than for $\xi=0$.  In this case a standard approach would be to compare the ratio of values of the \ac{PDF} of the data for $\xi\in\ac{HBR}$ and for $\xi$ unrestricted, i.e., to consider the ratio
$$
\max_{\xi\in\mathbb{R}^2} f(x;\xi)/ \max_{\xi\in\text{\ac{HBR}}} f(x;\xi),
$$
and it is easy to check that this leads to considering $W=\min_{\xi\in\text{\ac{HBR}}} \Delta(\xi)$.  Standard `large-sample'  arguments provide approximate inferences under which $W$ should be treated as a chi-square variable with one degree of freedom under the null hypothesis, but in the present situation these arguments do not apply unless the \ac{HBR} is large relative to $d_1$ and $d_2$, and it is usually better to use two degrees of freedom; see Appendix~\ref{app:dofquest}.  As $W=0$ when $x\in\ac{HBR}$, the corresponding significance probability then equals unity, giving the weakest possible evidence against the hypothesis that $\xi$ lies inside the \ac{HBR}, or equivalently the strongest possible evidence of a potential collision.  

Reference~\cite{Elkan22:StatFormCA} proposed a closely related approach to inference on the miss distance $\psi=\|\xi\|$. An exact pivot is unavailable to them, so they use approximate pivots, which depend only on the data and $\psi$, but whose precise distributions are unknown.  The two components of $\xi$ are expressed in terms of polar coordinates, $\psi$ and an angle $\lambda$, as  $(\xi_1,\xi_2) = \psi(\cos\lambda, \sin\lambda)$, and the nuisance parameter $\lambda$ is eliminated by maximization to reduce the problem to consideration of  $\psi$.  When $\psi$ equals the \ac{HBR}, the maximization is the same as was used to obtain $\min_{\xi\in\text{\ac{HBR}}} \Delta(\xi)$. This approach has the advantage of focusing on the miss distance, but the inference becomes approximate, though the authors of~\cite{Elkan22:StatFormCA} describe improved approximations that should be adequate in practice.   As with $\xi$, inferences on $\psi$ take either the form of a confidence set that contains the true $\psi$ with a specified probability, or a $P$-value that summarizes the evidence against a particular value of $\psi$, which would often be taken to be the \ac{HBR}. Since $\psi$ is a non-negative scalar, the corresponding confidence regions are intervals that may contain the value $\psi=0$. 

\begin{figure}[!t]
	\centering
	\includegraphics[width=1.0\linewidth,trim={0 8mm 0 0mm},clip]{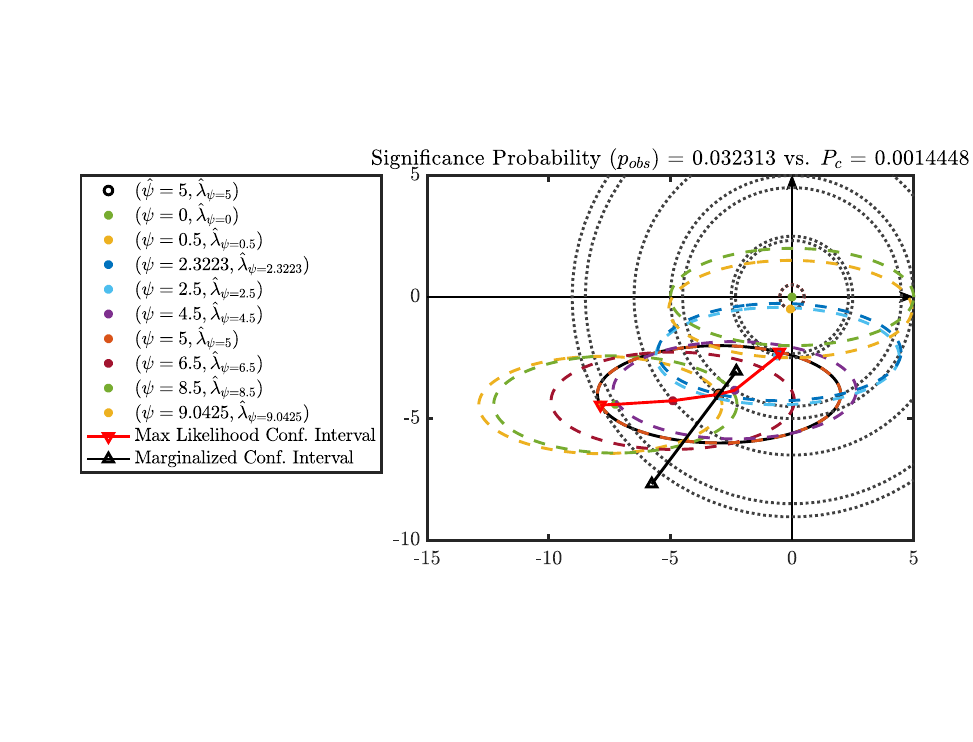}
	\caption{\label{fig:marginalizingCI} Illustration of the method of Reference~\cite{Elkan22:StatFormCA}. The solid ellipse centered at $\psi=5$ is the $1\sigma$ confidence region for the true position $\xi$ of the secondary object. The dotted circles represent a range grid, and the solid dots at each range are located at the clock angle that maximizes the joint likelihood over the nominal confidence region.}
\end{figure}

Figure~\ref{fig:marginalizingCI} illustrates the functioning of this algorithm. For comparison, the figure also shows a credible set based on marginalizing on the miss distance, and gives the corresponding value of $\hat{p}_c$.

The difference between the red and black \ac{CI}s in Figure~\ref{fig:marginalizingCI} is illuminating. The black credible set aligns with the predicted miss vector, along its radial component. This is a consequence of marginalizing the elliptical Gaussian \ac{PDF} shown by the ellipse along the miss vector direction to obtain a marginal \ac{PDF} for the miss distance. Reference~\cite{Carpenter:2019aa} showed that the use of $\hat{p}_c$ is equivalent to using the distribution of the miss distance to compute a credible set for it between zero and the \ac{HBR}. Such a set is complementary to a version of the black line shown in the figure with its lower and upper limits respectively at the \ac{HBR} and at infinity. While both the red and black sets correspond to the $1\sigma$ level, we see that, unlike the black set, the red set is fully contained within, and corresponds to the boundaries of, the $1\sigma$ confidence region; this coincidence arises because of the $\chi^2_2$ approximation for the likelihood ratio statistic. The red set is also a kind of a geodesic that stays as high as possible on the \ac{PDF} for a given distance from the origin, but the black set includes regions of much lower likelihood, particularly further from \ac{HBR}.

The Bayesian interpretation requires a prior \ac{PDF} for $\xi$.  If this \ac{PDF} is taken as uniform and the Gaussian model in Eq.~\ref{gaussian-pdf.eq} applies, then we saw in Section~\ref{sec:ca} that the posterior \ac{PDF} is the same formula.  In this case a highest posterior density credible region is a set that contains $\xi$ with a specified probability, but chosen so that no value of the posterior \ac{PDF} outside the region exceeds the smallest value within it.  Such a region is obviously elliptical, and can be defined as 
$$
\mathcal{H}_{1-\alpha} = \left\{ \xi: \int_{\xi\in\mathcal{E}} f(\xi\mid x)\,\D{\xi} =1-\alpha, \mbox{ where } \Delta(\xi)\leq c \mbox{ for some } c\geq 0\right\}.
$$
The credible region $\mathcal{H}_{1-\alpha}$ and the confidence region $\mathcal{E}_{1-\alpha}$ are identical, leading to the same inferences under either interpretation of the probability statements.

The probability dilution phenomenon that afflicts $\hat p_c$, whereby if $x\not\in \ac{HBR}$ then $\hat p_c$ tends to zero as either $d_1,d_2\to0$ or $d_1,d_2\to\infty$, with a maximum between them, does not apply to approaches based on $\Delta(\xi)$, which is monotonic decreasing in both $d_1$ and $d_2$.  

\subsection{Bases for Comparison}\label{sec:comparison}

Whatever assumptions are made, an operator making a decision about a conjunction faces four possible outcomes: (1) a decision to mitigate when the true relative position would otherwise have resulted in a collision (a correct decision); (2) a decision not to mitigate when the true relative position actually results in a miss (a correct decision); (3) a decision to do nothing when the true relative position actually results in a hit (an incorrect decision); and (4) a decision to mitigate when the true relative position would otherwise have resulted in a miss (an incorrect decision).  The last two of these outcomes, which represent two types of errors an operator can make, may be termed a missed detection and a false alarm. Over the course of many such decisions, any particular decision strategy can be characterized by the \ac{MDR} and \ac{FAR} for these opposing errors. Although it may seem that a decision strategy for satellite collision avoidance should enforce as small an \ac{MDR} as possible, regardless of the \ac{FAR}, this is impractical for several reasons, of which the most important but least understood is the low rate at which collisions are expected to occur.  Even in regimes such as low Earth orbit, where dozens of conjunctions occur each day, at the time of writing just one event (the 2009 collision between Iridium~33 and Cosmos~2251) can be classified as a missed detection decision error. Another reason is the relatively large uncertainty associated with conjunction predictions, which even less than a day before the \ac{TCA} typically enclose many times the volume of the \ac{HBR} sphere. Finally, false alarms are costly in terms of operational processes, lost mission productivity and satellite consumables, which might seem inconsequential for a single false alarm but mount rapidly when there might be hundreds of false alarms per year. 

Section~\ref{sec:infernc} described how standard assumptions enable exact statistical inference on the miss vector using the elliptical confidence/credible regions shown in Figure~\ref{fig:encpl}. However, because the uncertainties are typically large in comparison with the \ac{HBR}, unacceptably high \ac{FAR} would incur. Reference~\cite{Elkan22:StatFormCA} proposed an alternative based on approximate inference on the miss distance. This involves eliminating from the problem one or more nuisance parameters, not of ultimate interest and labeled $\lambda$, in order to focus discussion on interest parameters, labeled $\psi$, information about which is summarised in a pivotal quantity (or, more briefly, a pivot), i.e., a function of the data and $\psi$ whose distribution is known, at least approximately. For the short encounter \ac{CA} problem, viewing the miss vector in polar coordinates, the nuisance parameter $\lambda$ is the polar, or clock, angle, and the parameter of interest $\psi$ is the miss distance, or range.  To construct the pivots, \cite{Elkan22:StatFormCA} maximize their joint Gaussian likelihood by maximising over the clock angle at a fixed value of the range, and hence obtain a pivot whose normal distribution does not depend on $\lambda$, which may then be interpolated to find a value of $\psi$ corresponding to a desired probability.  One may then readily find either an interval on $\psi$ corresponding to a desired confidence level, or find the probability whose pivot corresponds to range equal to \ac{HBR}, which gives the significance probability, $p_{\rm obs}$, or $P$-value. Figure~\ref{fig:marginalizingCI} illustrates the functioning of this algorithm. For comparison, the figure also shows a confidence interval based on marginalizing on the miss distance, and a corresponding computation of $\hat{p}_c$.

\begin{figure}[!htb]
	\centering
	\includegraphics[width=0.99\linewidth]{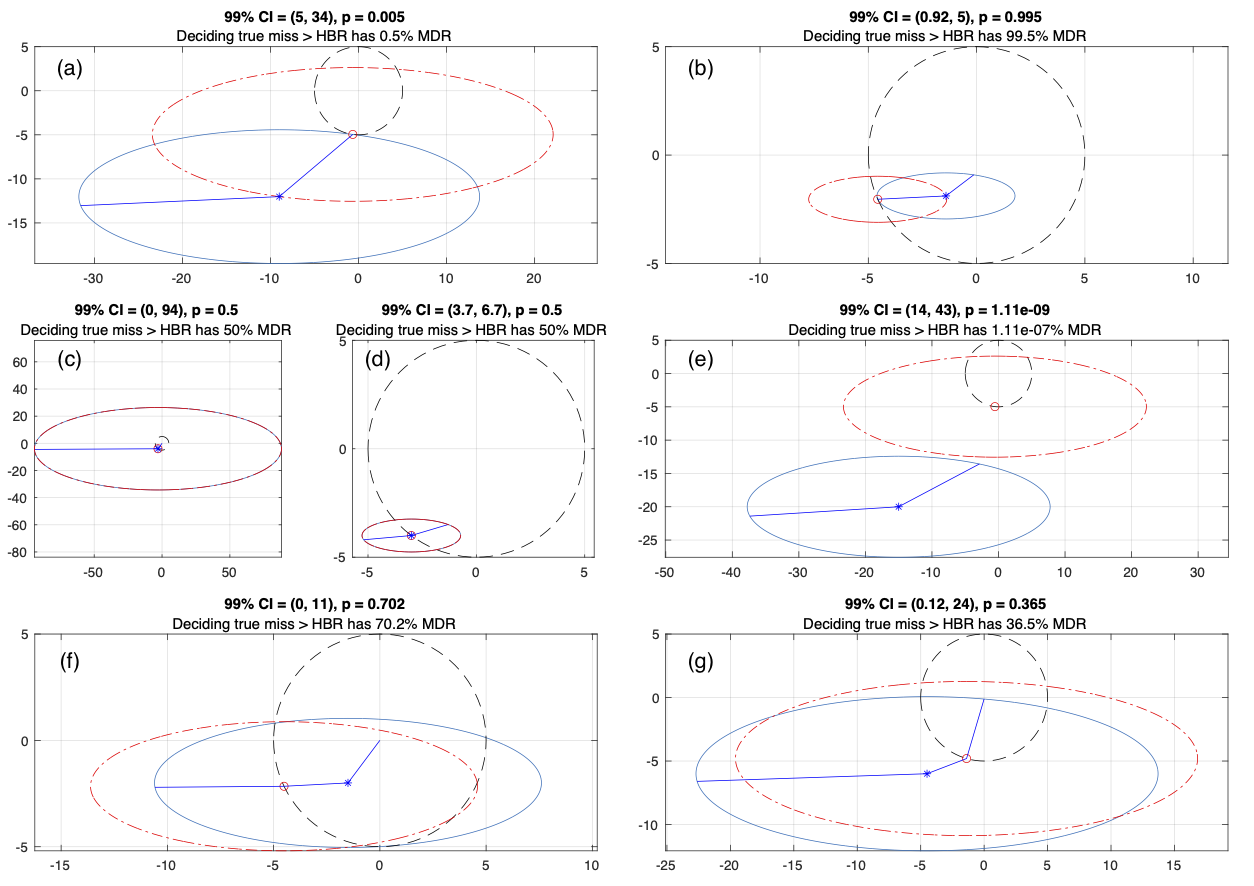}
	\caption{\label{fig:taxonomy1dof} Taxonomy of outcomes; in all subplots, the dashed black circle represents the combined hard-body region, the blue ellipse surrounding the blue asterisk is the nominal confidence region centered at the predicted miss vector, and the dash-dot red ellipse centered on the open red circle, provided for reference, is the nominal distribution centered on a maximum-likelihood miss vector with the radius of the \ac{HBR}. (a) Lower \ac{CI} limit = \ac{HBR}. (b) Upper \ac{CI} limit = \ac{HBR}. Miss Distance Estimate = \ac{HBR}, with (c) large uncertainty, and (d) small uncertainty. (e) Miss Distance Estimate $>>$ \ac{HBR}. (f) Miss Distance Estimate $<$ \ac{HBR}. (g) Miss Distance Estimate $>$ \ac{HBR}}
\end{figure}
It is illustrative to show a taxonomy of some of the main extreme cases that can occur when using the method of Reference~\cite{Elkan22:StatFormCA}. Figure~\ref{fig:taxonomy1dof} shows seven such cases.  For each case, a red open circle surrounded by a dash-dot red ellipse indicates the null hypothesis that the distribution is centered at the \ac{HBR}, at a clock angle that is consistent with the maximum likelihood of the blue distribution associated with the predicted miss vector, denoted by the blue asterisk.  As a reminder, the $P$-value is supposed to represent the probability that a prediction corresponding to the blue distribution, or a more extreme one, would be realized, if the red distribution corresponds to the truth. Thus, over many hypothetical repetitions of the same conditions, the \ac{MDR} of deciding the true miss distance was greater than the \ac{HBR} would be expected to converge to the $P$-value.

\begin{figure}[!htb]
	\centering
	\includegraphics[width=0.99\linewidth]{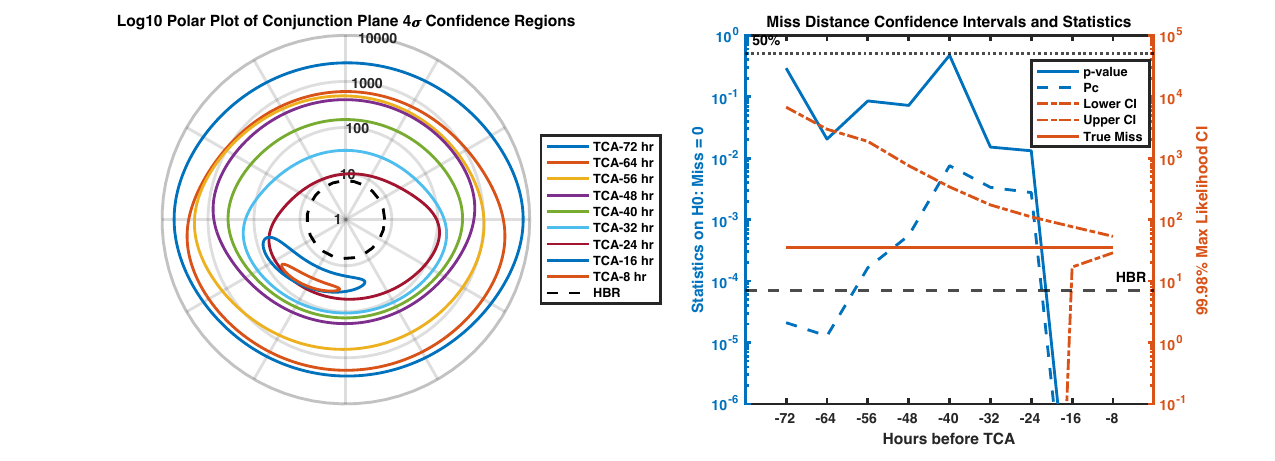}
	\caption{\label{fig:nearMissEx} Near miss example. The lower end of the confidence interval on the right is zero until 16~hours prior to \ac{TCA}, so does not appear on the plot until then.}
\end{figure}

\begin{figure}[!htb]
	\centering
	\includegraphics[width=0.99\linewidth]{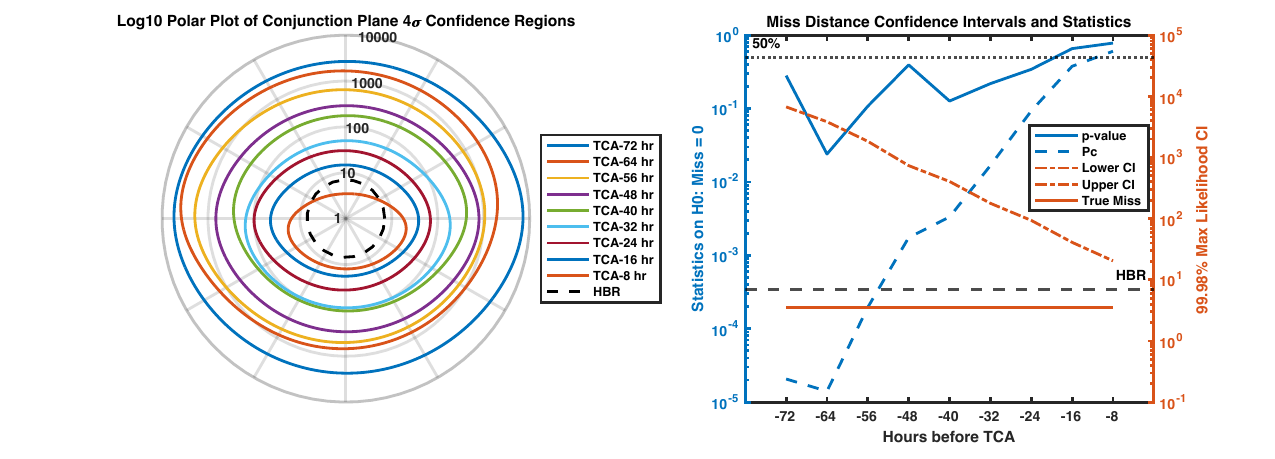}
	\caption{\label{fig:hitEx} Hit example. The lower end of the confidence interval on the right is zero throughout the time series, so never appears on the plot.}
\end{figure}

Figures~\ref{fig:nearMissEx} and~\ref{fig:hitEx} further illustrate the behavior of the method of Reference~\cite{Elkan22:StatFormCA}, and compare/contrast it to a strategy based on $\hat{p}_c$. Each figure plots the conjunction plane on the left, and a time series of miss-distance-related quantities on the right.  The conjunction plane uses polar coordinates, with a logarithmic range grid, which distorts the otherwise elliptical shapes of the time series of confidence regions leading up to \ac{TCA}, as the legend describes. The time series plot indicates the time evolution of the $P$-value and the $\hat{p}_c$ value using the left ordinate, and the evolution of a two-sided maximum-likelihood confidence interval for the miss distance using the right ordinate.
Figure~\ref{fig:nearMissEx} depicts a near miss.  It is notable that while both the $P$-value the $\hat{p}_c$ value decline rapidly once the uncertainty shrinks in the final 24 hours before \ac{TCA}, the $\hat{p}_c$ value had previously been quite low 3 days prior to \ac{TCA}.
Figure~\ref{fig:hitEx} depicts an illustrative case where the true miss distance was actually inside the \ac{HBR}. Interested readers may consult Reference~\cite{Elkan22:StatFormCA} for three more realistic case studies examining the performance and sensitivities of this approach in far greater detail.

While the approach as just described does not rely on the specification of any prior distribution for the true miss vector, Reference~\cite{Elkan22:StatFormCA} laid the foundation for the incorporation of such priors, and below we outline some further thoughts on this topic.

\section{Solution Approaches}\label{sec:apprchs}

\subsection{Decision Thresholds}
It has been common in \ac{CA} practice to compare $\hat{p}_c$ to a small threshold such as $1\times 10^{-3}$, $1\times 10^{-4}$, or even $1\times 10^{-5}$, for the purpose of deciding whether to begin a discussion about mitigation.  These guidelines have evolved over several decades, largely without regard to any kind of discussion about statistical inference and decision error rates, so it is  unsurprising that such thresholds may be inappropriate to a strategy about error rates.  Figures~\ref{fig:nearMissEx} and~\ref{fig:hitEx} make clear that, early in the timeline of a typical \ac{CA} process, the $P$-value may be orders of magnitude larger than $\hat{p}_c$, not atypically of the order of 10-50\%. One would be incorrect to infer from such values that the $P$-value is indicating anything like ``the probability of collision is 10-50\%.'' Rather, such values simply indicate that, due to the large uncertainties that are common early in the \ac{CA} timeline, the hypothesis that the true miss distance is small cannot be rejected. Although Figures~\ref{fig:nearMissEx} and~\ref{fig:hitEx} indicate that the $P$-value and $\hat{p}_c$ appropriately and similarly converge to either small or large values as the uncertainties shrink enough to make a clear decision possible, one can question whether decision thresholds associated with $\hat{p}_c$ should also be used with $P$-values. 

A potential advantage of extending the work of Reference~\cite{Elkan22:StatFormCA} to fully incorporate Bayesian priors is that this permits calibration of decision thresholds as targets for decision error rates, such \ac{MDR}, relative to the given prior.  For example, if the prior probability of a hit were $1\times 10^{-6}$, then a \ac{MDR} of $1\times 10^{-2}$ would result in odds of a hit being close to one in ten-thousand. Under the truncated uniform prior proposal of the next subsection, computation of the prior probability of a hit reduces to the trivial computation of the ratio of the area of the \ac{HBR} circle to the area of an elliptical slice through the screening volume. For a typical \ac{CA} problem, this area ratio is often of the order $1\times 10^{-6}$, which suggests a $P$-value threshold of the order of 1\% would be expected to reproduce the kinds of error rates that seem to be intended by a current practice that compares $\hat{p}_c$ to a threshold of $1\times 10^{-4}$. This discussion also suggests that if one prefers a \ac{CI} over a $P$-value as a decision indicator, something like a 99.5\% two-sided \ac{CI} would be far more appropriate than something like a 99.995\% two-sided \ac{CI}. 


\subsection{Truncated Priors}\label{trunc.sec}
\begin{figure}[!htb]
	\centering
	\includegraphics[width=0.99\linewidth]{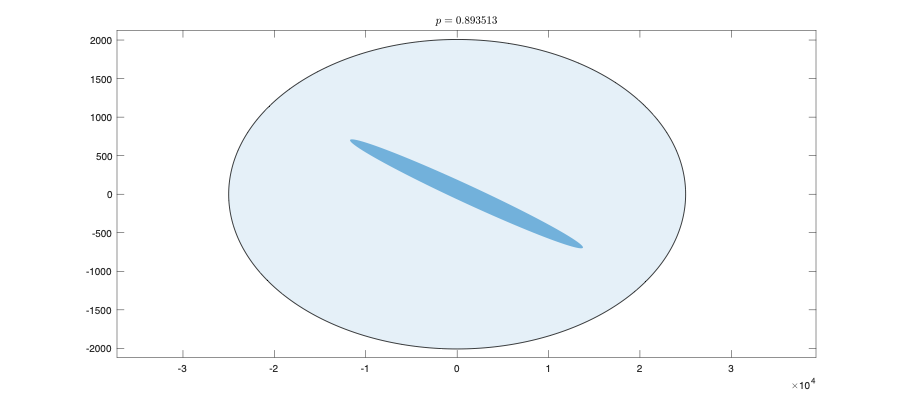}
	\caption{\label{fig:trunPriorInitial} Truncated prior corresponding to screening volume (light blue ellipse), with credible region corresponding to initial prediction (dark blue ellipse), at about a week prior to \ac{TCA}. Note that difference in axis scales.  The figure title shows a computation of the denominator of Eq.~\ref{eq:condensty}. }
\end{figure}

As mentioned in Section~\ref{sec:ca}, a uniform prior for the \ac{CA} problem seems unrealistic, and this may undercut the Bayesian interpretation of $p_c$. A possibly more justifiable prior is uniform in the screening volume and zero elsewhere.  For the short-encounter problem, this reduces to a truncated uniform prior in the conjunction plane, corresponding to the elliptical slice that the conjunction plane makes through the elliptical screening volume. For a typical screening volume, such a truncated prior might resemble the light blue ellipse in Figure~\ref{fig:trunPriorInitial}. A week or so prior to \ac{TCA}, the credible region associated with the prediction is often a sizable fraction of the screening volume and may even exceed it; the dark blue region in Figure~\ref{fig:trunPriorInitial} might be considered typical.

\begin{figure}[!htb]
	\centering
	\includegraphics[width=0.99\linewidth]{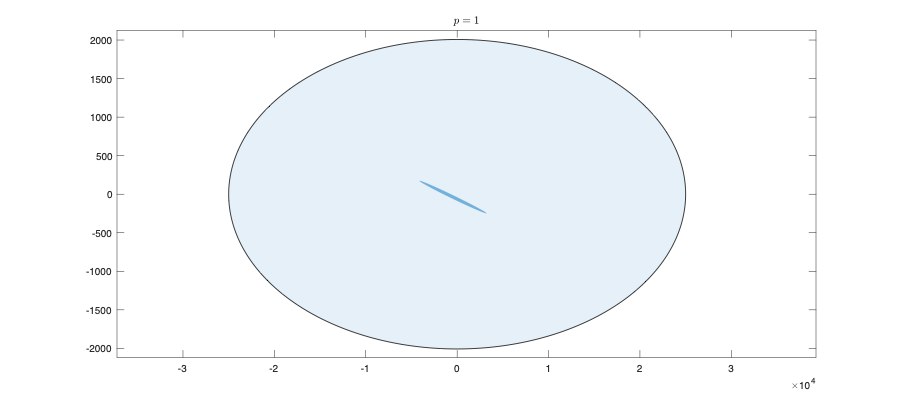}
	\caption{\label{fig:trunPrior3days} Truncated prior corresponding to screening volume (light blue ellipse), with credible region corresponding to initial prediction (dark blue ellipse), after a few days of tracking.  The figure title shows a computation of the denominator of Eq.~\ref{eq:condensty}.}
\end{figure}

After a few days of tracking, the credible region typically shrinks considerably: while it may still be much larger than the \ac{HBR}, it is usually a small fraction of the screening volume; see~Figure~\ref{fig:trunPrior3days}.

It is informative to compute the denominator of Eq.~\ref{eq:condensty}, $\int f(x\mid \xi)f(\xi)\,\d\xi$ for the situations shown in Figures~\ref{fig:trunPriorInitial} and~\ref{fig:trunPrior3days}.  The titles of each of these figures contain the results of such a calculation, computed using a method described by Das and Geisler\cite{das2024methodsintegratemultinormalscompute}. These results suggest that the error in assuming a uniform prior over all of space, when the actual prior might more realistically be described as truncated uniform prior of some kind, will usually be small, and will decrease as tracking reduces the size of the credible region closer to \ac{TCA}.

\subsection{Alternative Computation} \label{sec:altcomp}
It is common in \ac{CA} practice for poorly conditioned covariances to accompany miss vector predictions.  This primarily occurs because covariance predictions that do not fully account for prediction model uncertainty build-up over many days become highly extended in the direction of motion, resulting in large eigenvalue ratios. Such covariances can cause numerical problems for a variety of algorithms used in \ac{CA} practice, and the algorithm proposed in Reference~\cite{Elkan22:StatFormCA} is no exception. Perusal of Figures~\ref{fig:marginalizingCI} and~\ref{fig:taxonomy1dof} leads one to the following alternative, which requires a function for the distance of an ellipse from the origin in terms of the polar angle: Along a given $k\sigma$ elliptical contour, find the maximum and minimum distances from the origin; these give the $k\sigma$ \ac{CI} directly.
Next, vary $k$ such that the corresponding ellipse just touches the \ac{HBR}; that elliptical contour corresponds to the significance probability, $p_{\rm obs}$, which is the $P$-value.  Appendix~\ref{app:altimpmat} provides a \emph{Matlab} implementation of this alternative computation.

\subsection{Bayesian Extensions}\label{EB.sec}

Bayesian versions of any of the procedures described above require the specification of a prior probability distribution for the unknown parameters underlying the data. In the simplest setting in which 
$x = (x_{1}, x_{2})^{\T}$, which represents the position at which the secondary object will strike the conjunction plane, follows a bivariate normal distribution with mean and covariance matrix
$$
\mu = \begin{pmatrix} \psi \cos \lambda, \psi \sin \lambda \end{pmatrix}^{\T}, \quad
D^{-1} = \operatorname{diag}(d_1^2, d_2^2),
$$
this involves setting a joint prior distribution for the miss distance $\psi$ and the angle $\lambda$; as usual $D$ is considered to be known.  Taking independent uniform priors for $\psi$ and $\lambda$ or for the components of $\mu$ has difficulties discussed elsewhere \cite{Elkan22:StatFormCA}, and indeed any other choice might be queried on the grounds of being arbitrary.  Thus it might be better to take so-called empirical Bayes approach, whereby a prior is estimated from data on past conjunctions, each of which had parameter values drawn from the prior density.  To illustrate this, suppose that the joint prior density of $\lambda$ and $\phi=\psi^2$ is taken to be
\begin{equation}\label{prior.eq}
\dfrac{1}{2\pi}\times \dfrac{b^a}{\Gamma(a)} \phi^{a-1}e^{-b\phi}, \quad 0\leq \lambda< 2\pi, \phi>0, 
\end{equation}
i.e., $\lambda$ and $\phi$ are taken to be independent a priori with respective uniform and gamma \ac{PDF}s. In this formulation the parametric form of the density for $\phi$ is specified, but its parameters $a$ and $b$ are to be estimated from past conjunctions.  

Suppose now that conjunctions can be treated as independent, and that the $j$th of them has parameters $\lambda_j$, $\phi_j$ drawn from Equation~\eqref{prior.eq} and leads to observations $(x_{1j},x_{2j}, d_{1j}, d_{2j})$, with  $j=1,\ldots, n$ indexing the $n$ past conjunctions.  Then as
$
(x_{1j}/d_{1j}, x_{2j}/d_{2j})^{\T}
$
follows a bivariate normal distribution with mean vector 
$$
\tilde{\mu}_j = \begin{pmatrix} \dfrac{\psi_j \cos \lambda_j}{d_{1j}}, \dfrac{\psi_j \sin \lambda_j}{d_{2j} }\end{pmatrix}^{\T}
$$
and covariance given by the $2\times 2$  identity matrix, the sum of squares
\[
t_j = \left( \dfrac{x_{1j}}{d_{1j}} \right)^2 + \left( \dfrac{x_{2j}}{d_{2j}} \right)^2 
\]
follows a non-central chi-square distribution with two degrees of freedom and non-centrality parameter 
 \begin{equation}
\delta_j^2 = \|\tilde{\mu}_j\|^2 = \left( \dfrac{\psi_j \cos \lambda_j}{d_{1j}} \right)^2 + \left( \dfrac{\psi_j \sin \lambda_j}{d_{2j}} \right)^2 
= \dfrac{\phi_j}{2} \left\{ \left( \dfrac{1}{d_{1j}^2} + \dfrac{1}{d_{2j}^2} \right) + \left( \dfrac{1}{d_{1j}^2} - \dfrac{1}{d_{2j}^2} \right) \cos(2\lambda_j) \right\} .\label{noncentralityparam}
\end{equation}
This implies that the conditional mean and variance of $t_j$ are $\E(t_j\mid\lambda_j,\phi_j) = 2+\delta_j^2$ and $\var(t_j\mid \lambda_j,\phi_j) = 4(1+\delta_j^2)$, and thus the unconditional first and second moments of $t_j$ are
$$
\E(t_j) = 2 + \frac{1}{2} \E(\phi)A_j, \quad \E(t_j^2) = \E\{4(1+\delta_j^2) + (2+\delta_j)^2\} = 8 + 4\E(\phi) A_j + \frac{1}{4}\E(\phi^2)(A_j^2 + B_j^2/2), 
$$
where $A_j={1}/{d_{1j}^2} + {1}/{d_{2j}^2}$ and $B_j={1}/{d_{1j}^2} - {1}/{d_{2j}^2}$. 

We can now estimate the parameters $a$ and $b$ of the prior \ac{PDF} for $\phi$ using the empirical first and second moments of the $t_1,\ldots, t_n$.  If we write $\overline{A} = n^{-1} \sum_j A_j$, $\overline{A^2}=n^{-1}\sum_{j=1}^n A_j^2$ and $\overline{B^2}=n^{-1}\sum_{j=1}^n B_j^2$, then we see that 
$$
\E\left(\dfrac{1}{n}\sum_{j=1}^n t_j\right) =  2 + \frac12 \E(\phi)\overline{A}, \quad 
\E\left(\dfrac{1}{n}\sum_{j=1}^n t^2_j\right) =  8 + 4 \E(\phi) \overline A + \frac14\E(\phi^2)( \overline{A^2} + \overline{B^2}/2).
$$
Since $\E(\phi)=a/b$ and $\E(\phi^2) = a(a+1)/b^2$, equating $n^{-1} \sum t_j$ and $n^{-1}\sum t_j^2$ with their expectations and solving for $a$ and $b$ provides a data-based prior \ac{PDF} for $\phi$, and thus for the miss distance $\psi$.  This approach is simple, but the estimates may be numerically unstable if any of $d_{1j}$, $d_{2j}$ is small.  

Application of this method using the conjunction data described in the  following section led to $a \approx 0.23$ and $b\approx 0.0036$~km$^{-2}$, a gamma density for $\psi$ with a  pole at the origin and mean and standard deviation around 5.3~km and 6~km.  These surprisingly small values may stem from the numerical issues mentioned above.



\section{Results} 

To assess the classification performance of $\hat{p}_c$ and $p_{\rm obs}$ we construct a controlled dataset by selecting groups of at least 20 \ac{CDM}s corresponding to the same event. From each selected group, we extract the last recorded \ac{CDM}, which is the closest to the \ac{TCA}. The true relative position $\xi$ at this moment is taken as the baseline for generating both Hit and Miss cases. The dataset consists of 19,317 selected encounters, where for each encounter, we generate both a Hit case and a Miss case, ensuring a balanced dataset with $50\%$ Hits and $50\%$ Misses. 

A \textit{Hit} case is obtained by shifting the observed relative positions so that the objects collide, effectively placing the secondary object at the same location as the primary object at the \ac{TCA}. The final recorded covariance is then used to add artificial measurement uncertainties.
A \textit{Miss} case is based on interpolation between the original relative position and the Hit case using a shrinkage parameter $s$, using the formula
$$\xi_{\text{Miss}} = \xi_{\text{Hit}} + s (\xi_{\text{original}} - \xi_{\text{Hit}}).$$
This shifts a Miss towards a Hit while preserving the distinction between them. To ensure that the Miss remains outside the \ac{HBR}, we rescale the vector $\xi_{\text{Miss}}$ if its norm is smaller than the \ac{HBR} by multiplying by a rescaling factor 
$$
 \max\left(1, {\text{HBR}}/{\|\xi_{\text{Miss}}\|} \right).
$$
This guarantees that even after modification, the Miss remains a true non-collision scenario. 
 Larger values of $s$ keep the Miss cases farther from the Hits, making classification easier. As $s$ decreases, the Miss cases move closer to the Hits, reducing the separation between them and making classification more difficult. 
 
\begin{figure}[p]
    \centering 
    \begin{subfigure}[b]{0.45\textwidth}
        \centering
        \includegraphics[scale=0.4]{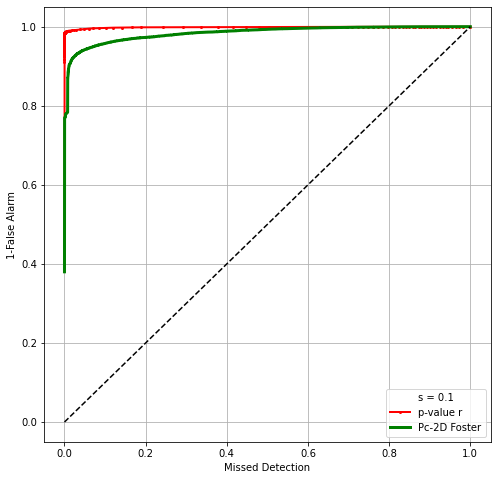}
        \caption{$s = 0.1$}
        \label{fig:roc_lambda_0_2}
    \end{subfigure}
    \begin{subfigure}[b]{0.45\textwidth}
        \centering
        \includegraphics[scale=0.4]{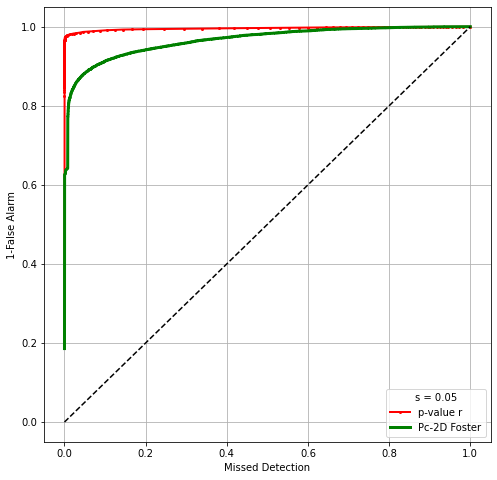}
        \caption{$s = 0.05$}
        \label{fig:roc_lambda_0_1}
    \end{subfigure}
    
    \vspace{0.5cm}
        \begin{subfigure}[b]{0.45\textwidth}
        \centering
        \includegraphics[scale=0.4]{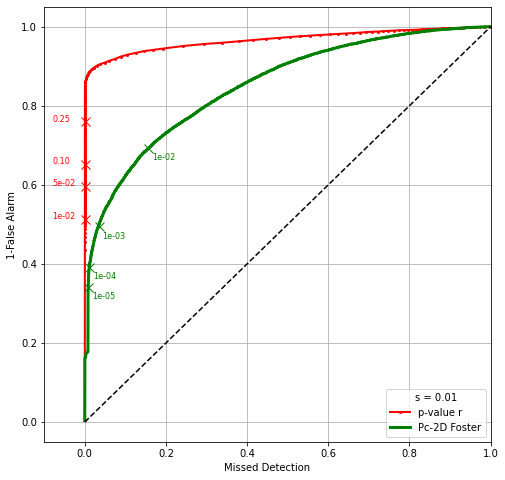}
        \caption{$s = 0.01$}
        \label{fig:roc_lambda_0_01}
    \end{subfigure}
    \begin{subfigure}[b]{0.45\textwidth}
        \centering
        \includegraphics[scale=0.4]{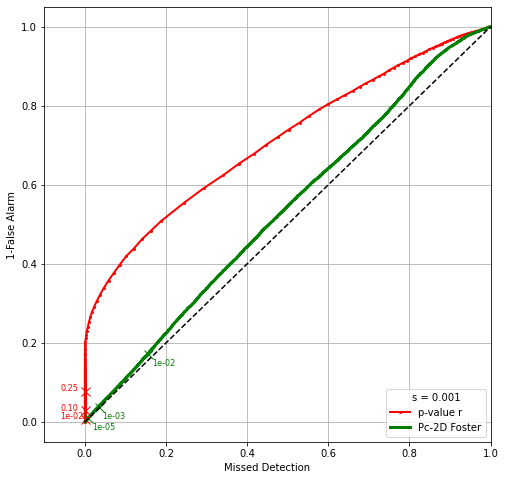}
        \caption{$s = 0.001$}
        \label{fig:roc_lambda_0_001}
    \end{subfigure}
    \caption{\ac{ROC} curves showing 1-False Alarm Rate (1-\ac{FAR}) as a function of the Miss Detection Rate (\ac{MDR}) for $p_{\rm obs}$  and $\hat{p}_c$ for different values of the separation parameter $s$.  The lower panels also show some of the values of $p_{\rm obs}$ (red crosses) and $\hat p_c$ (green crosses) used as thresholds in constructing the plots.}
    \label{fig:roc_curves}
\end{figure}

Figure~\ref{fig:roc_curves} shows \ac{ROC} plots that compare the performances of $\hat{p}_c$ (computed using the approach of~\cite{Fost92:collp}) and $p_{\rm obs}$ (using the likelihood root approach of~\cite{Elkan22:StatFormCA}) for different values of $s$. The $x$-axis represents the Miss  Detection Rate \ac{MDR}, i.e., the fraction of Hit cases that are incorrectly classified as Miss, while the $y$‑axis shows 1-False Alarm Rate (1‑\ac{FAR}), i.e., the fraction of Miss cases that are correctly classified as Miss. A perfect classifier would achieve an 1-FAR close to 1 together with a very low MDR, whereas a classifier performing at random would follow the diagonal $x=y$, so curves closer to the upper left corner of \ac{ROC} plots are preferable.
Taking collision as the null hypothesis might be confusing, so Table~\ref{tab:cadectab} shows how the terms used here are related to those used in conventional \ac{ROC} discussions.

\begin{table}[t]
    \centering
    \begin{tabular}{|c|c|c|c|}
    \hline
\multicolumn{2}{|c|}{\multirow{2}{*}{Conjunction Assessment}}  &   \multicolumn{2}{c|}{Null hypothesis (Hit) is}  \\
    \cline{3-4}
\multicolumn{2}{|c|}{}                    &   True (Hit)    &   False (Miss)       \\
    \hline
\multirow{2}{*}{\makecell{Decision \\ about null \\ hypothesis \\
(Hit) is}}      &   \makecell{Don't reject: \\ Mitigate \\ (Maneuver)}          &  \makecell{Correct inference \\
(true negative) \\
(probability = $1-\alpha$)\\ \textbf{Detection}}     &   \makecell{Type II error \\
(false negative) \\
(probability = $\beta$) \\ \textbf{False Alarm}}           \\
\cline{2-4}
                        &   \makecell{Reject: \\ Dismiss \\ (No Maneuver)}         &   \makecell{Type I error \\
(false positive) \\
(probability =  $\alpha$) \\ \textbf{Missed Detection}}       &   \makecell{Correct inference \\
(true positive) \\
(probability = $1-\beta$)}           \\
    \hline
\end{tabular}
    \caption{Conjunction Assessment Decision Table}
    \label{tab:cadectab}
\end{table}

We construct the curve for $p_{\rm obs}$ using a uniform grid of thresholds in the interval $[0,1]$, which is naturally spanned by the $P$-values. Collision-probability estimates $\hat{p}_c$ tend to be much smaller, so we construct its curve using thresholds equal to observed $\hat{p}_c$ values. This ensures that every distinct collision-probability cutoff is tested when computing the \ac{ROC} curves.  Some of the threshold values used are shown in the lower panels. 

The lower left region of these graphs corresponds to extremely low thresholds, where both the $p_{\rm obs}$ and $\hat{p}_c$ values are very small. As a result, almost all Hit cases (which ideally should yield high $p_{\rm obs}$) are correctly classified as Hits, leading to a very low miss detection rate \ac{MDR}. However, nearly all Miss cases also exceed this minimal threshold, resulting in a very high false alarm rate (and thus a very low 1-\ac{FAR}). In other words, with an overly low threshold, both Hits and Misses are almost uniformly classified as Hits.
Conversely, in the upper right region, where the threshold is set very high, the $p_{\rm obs}$ and $\hat{p}_c$ values are less extreme. Here, many Hit cases fall below the high threshold and are misclassified as Misses, yielding a high \ac{MDR}, even though most Miss cases are correctly classified (resulting in a high 1-\ac{FAR}). This overly conservative threshold minimizes false alarms at the expense of missing many true collisions. Together, these extremes highlight the importance of selecting an appropriate threshold that balances missed detections against false alarms to achieve optimal classification performance.

Figures~\ref{fig:roc_lambda_0_2}--\ref{fig:roc_lambda_0_1} show that at larger values of $s$, where the Hits and Misses are well-separated, both $\hat{p}_c$ and $p_{\rm obs}$ perform satisfactorily. However, as $s$ decreases,  classification becomes more challenging, and the detection rate of $p_{\rm obs}$  for a given false positive rate is uniformly higher than that of $\hat{p}_c$; see Figures~\ref{fig:roc_lambda_0_01}--\ref{fig:roc_lambda_0_001}.  Although these results reflect controlled, synthetic scenarios and may not translate directly to real‐world operational performance, they illustrate how the methods separate collision (Hit) and non‐collision (Miss) cases under varying levels of class overlap. 
 
 
At first sight it may seem surprising that $p_{\rm obs}$ uniformly dominates $\hat p_c$, but a little reflection shows that the converse would be more striking.  The Neyman--Pearson lemma implies that a likelihood ratio test is most powerful for the comparison of two simple hypotheses, and although the hypotheses here are composite, the lemma nevertheless suggests that the likelihood ratio test will be difficult to beat.  The lemma also suggests that the \ac{ROC} curve of $p_{\rm obs}$ dominates that of $\hat p_c$ even when the two classes are well-separated and thus the differences between the two approaches are less consequential. 

\section{Conclusions}
The `collision probability' $p_c$ generally employed in current satellite collision risk assessment presents the practitioner with a number of problems.  It does not represent the probability that the true miss distance will be smaller than the proximity threshold (\ac{HBR}) but rather that the uncertain (noisy) estimate of the miss will be smaller than this threshold --- not a calculation directly relevant to assessing the likelihood of an actual collision.  In fact, it is not a frequentist probability at all but (tacitly, since this aspect of the calculation is rarely directly discussed) a Bayesian construct with an implicit uniform prior --- which is not a compelling choice, because the process of identifying a close approach substantially restricts the possible set of miss distances.  The calculation additionally suffers from ``false confidence'' in that situations of small true miss distance but large uncertainty can produce unexpectedly high missed detection rates when used with typical $p_c$ mitigation thresholds.  These issues compound to present the decision-maker with a quantity that does not have the properties that he or she has been led to expect.

Two different approaches have been proposed to improve the situation.  The first, originally developed in \cite{Elkan22:StatFormCA}, reframes the problem in frequentist terms:  the true miss distance is treated not as a distribution but as an unknown constant, and an inferential construct is assembled to produce a confidence interval, and/or calculate a p-value, that the miss vector estimation process would produce the observed results under the null hypothesis that the true miss distance is equal to the \ac{HBR}.  Very small significance levels indicate the improbability of the estimate of so large a normalized miss given the presence of an actual hit, and thus they would counsel the rejection of the null hypothesis that the conjunction represents a  collision.  The confidence interval construction process for this approach was treated in detail, with a number of edge cases explored.  The second method returns to the Bayesian framework but attempts to improve the prior, with two different approaches suggested.  The first is to employ a truncated uniform prior, related to the volume of space examined to find conjunctions initially.  While more firmly justified than the uniform prior, its actual effect on the $p_c$ calculation is in most cases very small.  The second approach is to use data from previous (presumed independent) miss vector estimates for this same event, fit to a gamma distribution functional form.  Comparative performance of the frequentist construct to the traditional $p_c$ calculation, using a synthetically altered dataset, suggests that the frequentist approach can be preferable in certain situations.

While these alternatives to the currently-calculated $p_c$ appear promising, their ultimate suitability for operational use will hinge on the ability to manage both their missed detection and false alarm rates.  As pointed out in Section~\ref{sec:comparison}, control of the false alarm rate (counterintuitively) is often the greater consideration, given the very low probability of an actual collision and the often severe degree of mission disruption and lifetime reduction of frequent collision risk mitigation maneuvers.  A profiling campaign is needed to compare these alternatives to the $p_c$ using a very large database of historical conjunction information to characterize the magnitude of the increased false alarms, which is expected on theoretical grounds, and to determine whether their rate can be reduced to operationally acceptable levels while preserving a meaningful test statistic.  It is hoped that such control can be achieved through judicious setting of the significance thresholds and cognizance of external considerations, such as the severity of the space debris production that a particular conjunction might be expected to manifest should it result in a collision.

\section*{Acknowledgements}

 This material is based upon work partially supported by the Air Force Office of Scientific Research under award number FA8655-24-1-7009

\begin{appendices}
\section{Sensitivity to Covariance Realism}
\label{app:covrealsens}
\begin{figure}[!hb]
    \centering
    \includegraphics[width=0.99\linewidth]{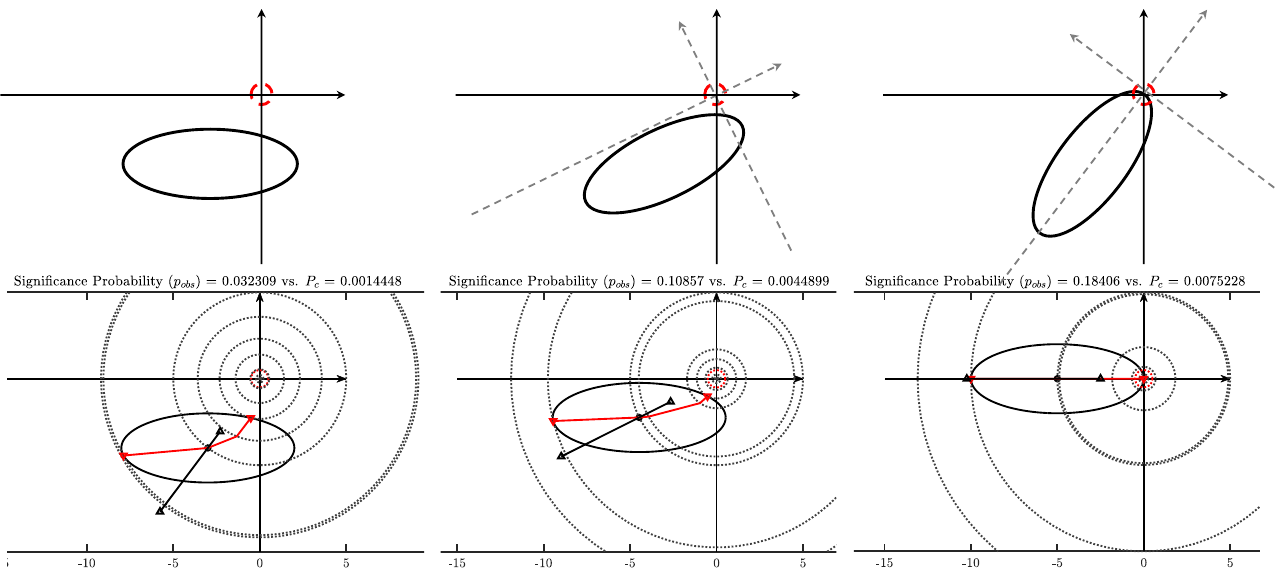}
    \caption{Sensitivity to Rotation of Confidence/Credible Region}
    \label{fig:rotSens}
\end{figure}

An ongoing concern in the broader \ac{CA} community of practice is that the covariance matrices provided either by space traffic monitoring organizations, or the owner/operators of active satellites themselves, may not adequately reflect the degree of variability that might realistically be expected due to the effects of errors in the tracking data, modeling uncertainties, random events such as arise from space weather, etc. Often, the providers of such information want to include additional conservatism in their products, which compounds the already difficult problem that much of the time, uncertainties are simply too large to make effective decisions, as Reference~\cite{Carpenter:2019aa} describes. While one might  expect that covariances that are either too large or too small will similarly affect both the $\hat{p}_c$ and the $P$-value, this work has already shown that the $P$-value is far less sensitive to underestimating the risk when the covariances are large early in the \ac{CA} timeline, which suggests the $P$-value would also be less sensitive to understating the risk when covariances are conservatively inflated at any point in the timeline. What is less clear however is whether or not the $P$-value is more or less sensitive to errors in orientation of the elliptical confidence region derived from the covariance. As Section~\ref{sec:altcomp} has already described, a common issue with longer predictions is that such elliptical regions become highly elongated, with the result that a small change in the orientation of the ellipse within the conjunction plane can dramatically affect the likelihood in the neighborhood of the \ac{HBR} circle. Figure~\ref{fig:rotSens} illustrates this sensitivity for both the $\hat{p}_c$ and the $P$-value, as well as the marginalized and maximum likelihood confidence intervals, for three different rotations of the same confidence region at the same predicted distance from the origin. As the calculations shown in the middle of each panel show, the relative change in the $\hat{p}_c$ and the $P$-value across these variations in rotation angle are quite similar, indicating there is no significantly greater sensitivity to such variations for one or the other risk metric.

\section{Degrees of Freedom for Likelihood Ratio Statistic}
\label{app:dofquest}
The large-sample distribution of the likelihood ratio statistic for comparison of two nested hypotheses in a so-called regular case is chi-square with degrees of freedom given by the difference in the numbers of parameters fixed under the two hypotheses.  In the present setting this approximation may fail.  To see why, suppose that $x=(x_1,x_2)$ is the observed position in the conjunction plane of the secondary object relative to the first, and let $\xi=(\xi_1,\xi_2)$ denote its true position.  Suppose for simplicity that $x\sim \N_2(\xi, d^2I_2)$, so the measurement error is circularly symmetric.  The likelihood ratio statistic for testing the hypothesis that $\|\xi\|=\psi_0$ against the alternative that $\|\xi\|>\psi_0$ is 
$$
W = d^{-2}\left\{ (x_1-\hat\xi_1)^2 + (x_2-\hat\xi_2)^2\right\}, 
$$
where $\hat\xi_1^2 +\hat\xi_2^2 = \psi_0^2$, and the circular symmetry yields $\hat\xi_1 = \psi_0 x_1/R$ and $\hat\xi_2 = \psi_0 x_2/R$, where $R^2=x_1^2+x_2^2$.  It is then straightforward to check that $W = (R-\psi_0)^2/d^2$, and if we then write $x_1=\xi_1+d \v_1$, $x_2=\xi_2+d\v_2$, where $\v_1,\v_2$ are independent standard normal variables, and under the null hypothesis let $\delta^2=\|\xi\|^2/d^2 = \psi_0^2/d^2$, we obtain
$$
W \Deq \left[\{(\delta+\v_1)^2 + \v_2^2\}^{1/2}-\delta\right]^2
$$
where $\Deq$ is to be read as `has the same distribution as'; the quantity in the braces has a non-central chi-square distribution with two degrees of freedom and non-centrality parameter $\delta^2$. If $\delta\to 0$, i.e., the testing radius $\psi_0=\|\xi\|$ becomes very small compared to $d$, then the distribution of $W$ is approximately that of $\v_1^2+\v_2^2$, i.e., $\chi^2_2$.  This will often be the case, as measurement uncertainty will typically be very large compared to the radius $\psi_0$ to be tested.   Heuristically we see that if uncertainty increases without limit, then the null hypothesis $\|\xi\|=\psi_0$ is barely different from testing the point hypothesis $\|\xi\|=0$, which would give $W\sim \chi_2^2$.  If by contrast $\delta\to\infty$, i.e., the uncertainty becomes small relative to the testing radius, then 
$$
W \Deq \delta^2\left[\{1+ 2\v_1/\delta + (\v_1^2+\v_2^2)/\delta^2\}^{1/2} - 1\right]^2 = \left\{ \v_1 + (\v_2^2-\v_1^2)/(2\delta) + O(\delta^{-2})\right\}^2,
$$
and as $\delta\to\infty$ the limiting term is $\v_1^2 \sim \chi^2_1$, corresponding to the conventional large-sample approximation, which corresponds to the variance of $x$ shrinking to zero.  A similar but more involved computation applies when the distribution of $x$ is elliptical rather than circular in shape.  If it is unclear which of the two approximate distributions should be used in a given situation, Monte Carlo simulation of the exact distribution of $W$ for given $\psi_0$ and error variances $d_1^2$ and $d_2^2$ would be rapid, but this would  be needed  only when the lower and upper bounds to the value of $p_{\rm obs}$ obtained using the $\chi_1^2$ and $\chi^2_2$ approximations for $W$ give contradictory evidence about the likelihood of collision.

\begin{figure}[!ht]
    \centering
    \includegraphics[width=0.7\linewidth]{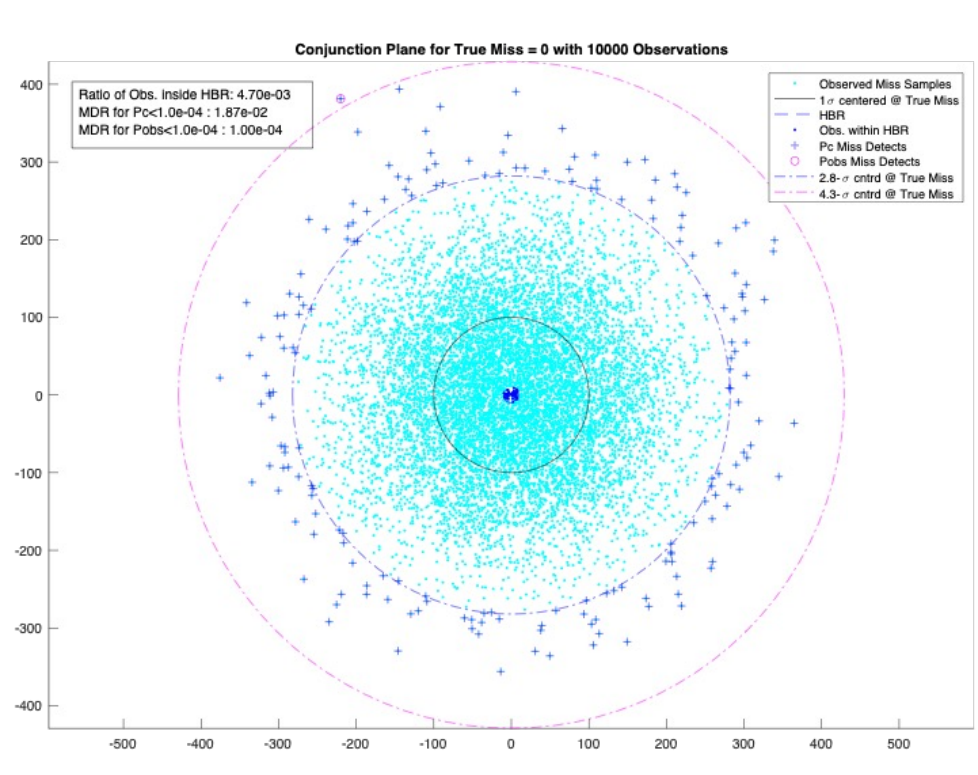}
    \caption{$n_{\text{DoF}} = 2$, \ac{MDR} $\sim\alpha$}
    \label{fig:mdr2dof}
\end{figure}

In Figure~\ref{fig:mdr2dof} the threshold for the $P$-value was set using a $\chi^2$ distribution with two degrees of freedom, whereas in Figure~\ref{fig:mdr2dof} the threshold for the $P$-value was set using a $\chi^2$ distribution with one degree of freedom. Everything else, except the particular set of 10,000 realizations, is the same across both figures. The threshold was intended to target a \ac{MDR} of $1\times 10^{-4}$. These results suggest the correct degrees of freedom here is two.  
\begin{figure}[!ht]
    \centering
    \includegraphics[width=0.7\linewidth]{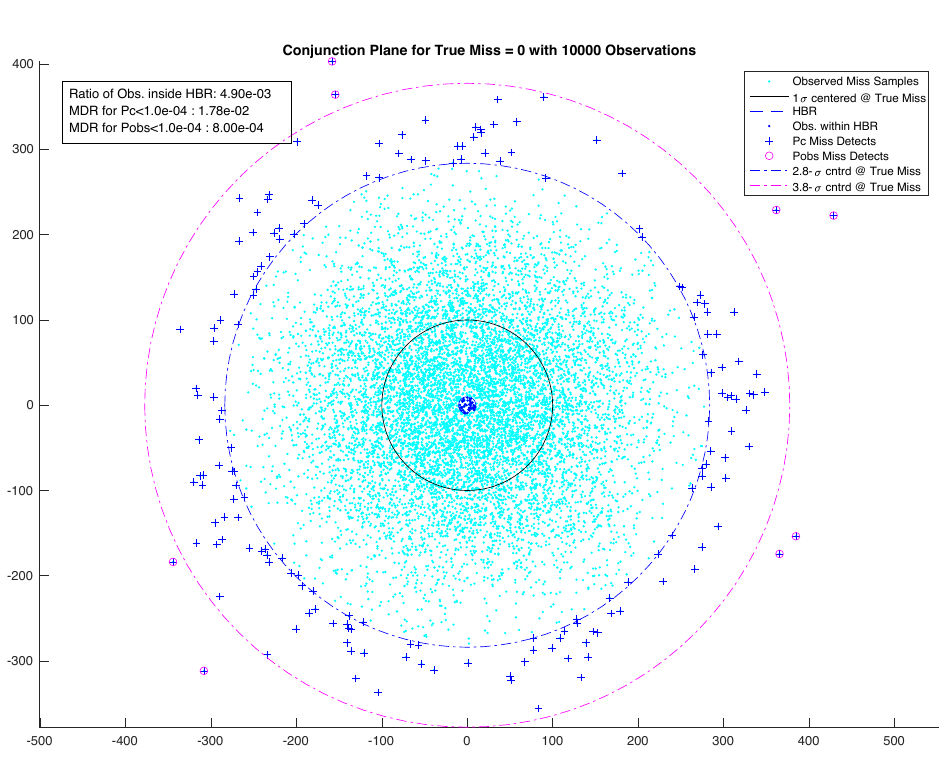}
    \caption{$n_{\text{DoF}} = 1$, \ac{MDR} $10\sim\alpha$}
    \label{fig:mdr1dof}
\end{figure}

\section{Alternative Implementation \emph{Matlab} Function }
\label{app:altimpmat}
\begin{lstlisting}[style=Matlab-editor]
function [ci,pval,z] = mlcipval2d(x,P,HBR,alph,ndof)
%MLCIPVAL2D Max likelihood CI & p-value for 2D CA
%   Alternative implementation of method of Elkantassi & Davison JGCD 45(12)
arguments
    x (2,1)
    P (2,2)
    HBR (1,1)
    alph (1,1)
    ndof (1,1)
end

%% Confidence interval
%z_1mao2 = -norminv(alph/2); % by symmetry is equal to norminv(1-alph/2);
z_1mao2 = sqrt(chi2inv(1-alph,ndof)); % same as above if ndof=1
dmin = ksig2origin(x,P,z_1mao2,"min");
dmax = ksig2origin(x,P,z_1mao2,"max");
if sqrt(x'/P*x) <= z_1mao2 % does z_1mao2 ellipse contain the origin?
   dmin = 0;
end
ci = [dmin,dmax];
z = [z_1mao2 z_1mao2];

%% Significance probability
% Want z_p < 0 when deriving from min dist and vice versa.  Handle trivial
% cases first to avoid needless computation.
dx = norm(x);
kmax = 7;
if dmin == HBR
    z_p = -z_1mao2; % Ps <= 1/2
elseif dmax == HBR
    z_p = z_1mao2; % Ps >= 1/2
elseif dx == HBR
    z_p = 0; % Ps = 1/2, no matter size of P
elseif ksig2origin(x,P,kmax,"min") > HBR % far away, p = 0
    z_p = -Inf;
elseif dx < HBR % stretch or shrink ellipse s/t max dist = HBR
    z_p = fzero(@(z) HBR - ksig2origin(x,P,z,"max"),[0 kmax]);
else % stretch or shrink ellipse s/t min dist = HBR
    z_p = -fzero(@(z) HBR - ksig2origin(x,P,z,"min"),[0 kmax]);
end
%pval = normcdf(z_p); % same as below if ndof=1
if z_p <= 0
    pval = (1-chi2cdf(z_p^2,ndof))/2;
else
    pval = (1+chi2cdf(z_p^2,ndof))/2;
end
z = [z z_p];

end

%% Subfunction
function dopt = ksig2origin(x,P,k,maxmin)
if nargin == 3
    maxmin = 'min';
end
% There may be a more direct way to compute d(th) than below, e.g. 
% http://www.jaschwartz.net/journal/offset-ellipse-polar-form.html
[V,lambda] = eig(P,'vector');
a = lambda(1);
b = lambda(2);
m = @(th) sqrt((a*b)./(a*sin(th).^2 + b*cos(th).^2));
u = @(th) k*m(th).*cos(th);
v = @(th) k*m(th).*sin(th);
d = @(th) vecnorm(V*[u(th); v(th)] + x);
if strncmpi(maxmin,'max',3)
    [~,dopt] = fminbnd(@(th)-d(th),0,2*pi);
    dopt = -dopt;
else
    if x'/P*x <= k^2 % is origin inside ksig ellipse?
        dopt = 0;
    else
        [~,dopt] = fminbnd(d,0,2*pi);
    end
end
end
\end{lstlisting}

\end{appendices}
\clearpage
\bibliographystyle{unsrt}
\bibliography{SpOps25_InfMissCA}

\end{document}